\documentclass[%
 reprint,
 amsmath,amssymb,
 aps,
]{revtex4-2}

\usepackage{graphicx}
\usepackage{dcolumn}
\usepackage{bm}
\usepackage[dvipsnames]{xcolor}
\usepackage{lineno}
\usepackage[normalem]{ulem}
\newcommand{\ag}[1]{\textcolor{red}{#1}}

\begin{document}

\preprint{APS/123-QED}

\title{Non-Hermitian zero mode laser in a nanophotonic trimer}

\author{Kaiwen Ji}

\author{Bruno Garbin}

\author{Melissa Hedir}

\author{Juan A. Levenson}

\author{Alejandro Yacomotti}%
\email{alejandro.gacomotti@c2n.upsaclay.fr}
 
\altaffiliation[$^1$]{Centre de Nanosciences et de Nanotechnologies, CNRS, Université Paris-Sud, Université Paris-Saclay, 10 Boulevard Thomas Gobert, 91120 Palaiseau, France}

\date{\today}

\begin{abstract}
Symmetry-protected zero modes in arrays of coupled optical elements have attracted considerable attention because they are expected to be robust against coupling disorders. In the Hermitian limit, zero modes are dark ones, i.e. the intensity in one sublattice vanishes; yet, in a non-Hermitian counterpart, zero modes can be bright and feature $\pi/2$ phase difference between sublattices. In this work, we report on the direct observation of a lasing zero mode in a non-Hermitian three coupled nanocavity array. We show efficient excitation for nearly equal pump power in the two extreme cavities. Furthermore, its efficiency can be dynamically controlled by pumping the center cavity. The realization of zero mode lasing in large arrays of coupled nanolasers has potential applications in laser-mode engineering and it opens up promising avenues in optical computing.
\end{abstract}

\maketitle


Majorana zero modes are collective excitations pinned at the middle of a gapped bandstructure. Their topological and non-Abelian properties make them immune against certain types of disorder, and therefore robust for applications in topological quantum computing \cite{RevModPhys.80.1083}. In optics, they have been experimentally demonstrated in a few different platforms, such as flat band structures in optical waveguide arrays~\cite{vicencio2015observation,mukherjee2015observation} and topologically protected gap modes in a one dimensional Su-Schrieffer-Heeger (SSH) chain~\cite{st2017lasing,parto2018edge,han2019lasing}. The zero modes originate, in these experiments, from the chiral --or sublattice-- symmetry, $\{H,C\}=0$, where $H$ is the Hamiltonian and $C$ a unitary operator. On the other hand, zero modes have been observed in condensed matter physics and topological superconductors~\cite{sun2016majorana,lutchyn2018majorana,jack2019observation}; they result from particle-hole symmetry 
(PH, also known as charge-conjugation symmetry), where the Hamiltonian anti-commutes with an anti-unitary operator $CT$ ($\{H,CT\}=0$, where $T$ is the time reversal operator).

In the Hermitian limit, both chiral and particle-hole symmetries ensure that the eigenvalues always appear in pairs, $\epsilon_i=-\epsilon_j$~\cite{ge2017symmetry}. Such a symmetrical band leads to a zero mode with $\epsilon=0$ for $i=j$. However, in the non-Hermitian realm, the eigenvalues are generally complex and the imaginary parts account for loss/gain rates. Consequently, the chiral and particle-hole symmetries generally result in different eigenvalue spectra, i.e., $\epsilon_i=-\epsilon_j$ still holds for the chiral symmetry case~\cite{rivero2021chiral}, while $\epsilon_i=-\epsilon^*_j$ takes place in systems with non-Hermitian particle-hole (NHPH) symmetry~\cite{pikulin2012topological}, where the $^*$ is the complex conjugate. Therefore, the zero mode features $\Re[\epsilon]=0$ for the NHPH symmetry, meaning that the non-Hermitian zero mode is more robust than its Hermitian counterpart, since  no restriction is applied to its imaginary part. Recently, NHPH-symmetry protected zero mode has been demonstrated in photonic systems such as a PT symmetric
waveguide array with defect and topological segment~\cite{qi2018defect,pan2018photonic}. However, in those approaches, only one zero mode can be realized at a given spatial location, for instance, the defect or the topological boundary. On the other hand, zero modes warranted by NHPH-symmetry in a three coupled photonic crystal cavity array have been reported recently, but single-spot pumping conditions prevented efficient excitation and therefore no lasing zero mode has been observed ~\cite{hentinger2022direct}.

In this work, we report on the first direct observation of a lasing zero mode warranted by NHPH-symmetry in a three coupled nanocavity array with embedded quantum wells (QWs). The lasing condition is enabled by spatially patterning the pump spot by means of a spatial light modulator (SLM). Such a prototype can be easily extended to a two-dimensional network supporting an arbitrary number of zero modes with different intensity distributions~\cite{ge2017symmetry}. In our three cavity case, at least one zero mode is expected to exist at the frequency of the single cavity because the number of cavities is odd~\cite{ge2017symmetry}, and it can be efficiently excited above laser threshold by pumping the two extreme cavities with similar optical powers. Furthermore, we dynamically control the emission intensity by varying the pump power in the center cavity. 

\begin{figure}[t!]
	\centering
	\includegraphics[width=\linewidth]{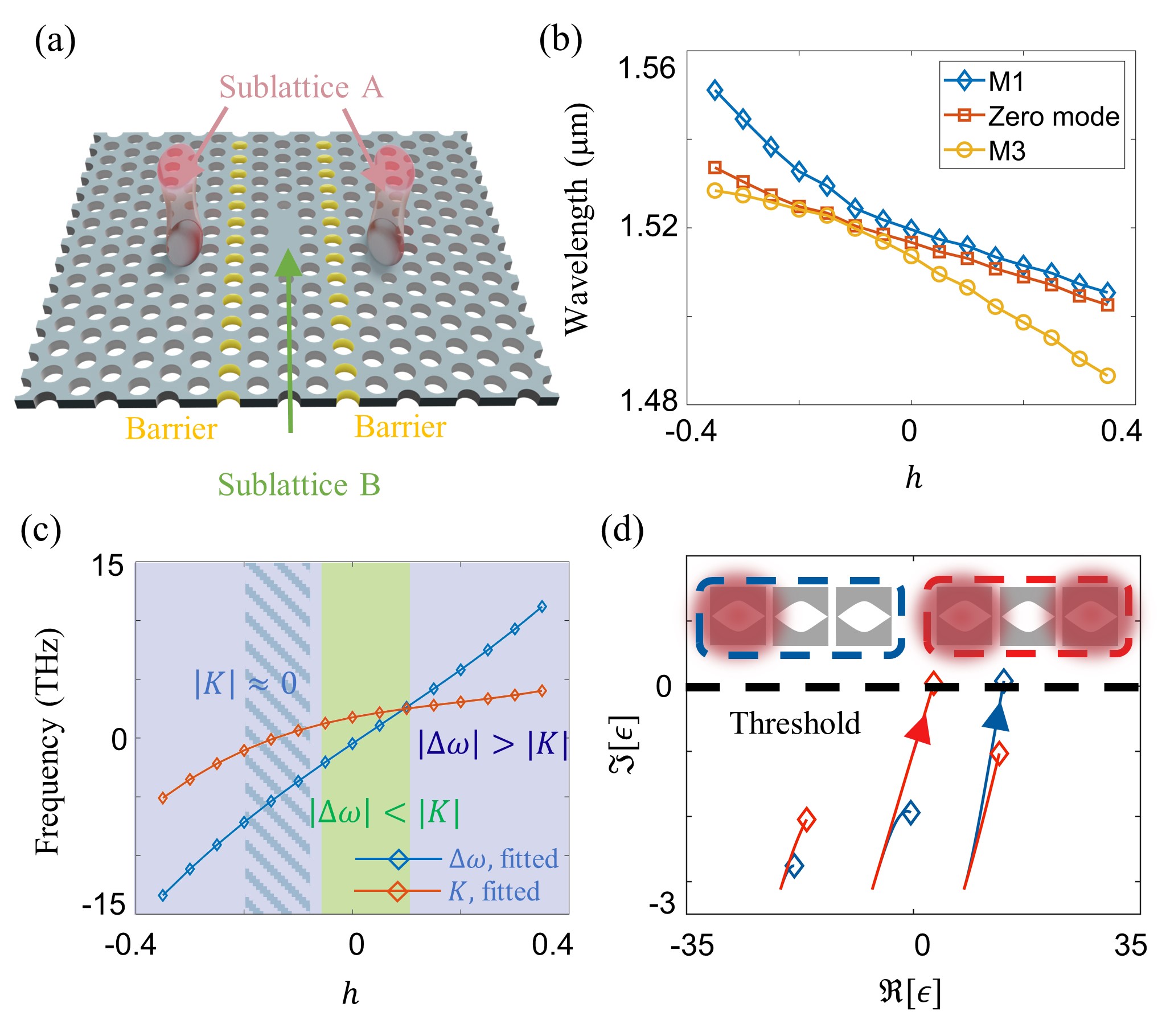}
	\caption{(a) Schematic of the three coupled cavity system. The yellow arrays represent the barriers. (b) FDTD simulations showing the coupling and the detuning as a function of barrier size $h$. (c) Coupling and cavity detuning extracted from the hybrid mode frequency splitting and average of two coupled PhC cavities’ 3D FDTD numerical simulation. Solid lines are third-order polynomial fittings, which are subsequently used for CMT calculation with three coupled cavities in the main text. (d) Evolution of eigenvalues, the arrows indicate the direction of trajectories, the black dashed line represents the laser threshold $\Im[\epsilon]=0$. Here, $P$ increase from $0$ to $P_0$ and $1.45P_0$ (where $P_0$ is the threshold of a single cavity) for pumping two extreme cavities (blue curves) and pumping a single cavity (red curves), respectively. The diamonds represent the end point.  Here the parameters are: $\tau=7.1$ps, $\beta=0.017, \alpha=3,  \Gamma_\parallel=2.2$GHz, $\Gamma_{tot}=5$GHz, $K=10/\tau$, $n_0=10^{18}$cm$^{-3}\times V_a$, with $V_a=0.016\times 10^{-12}$cm$^3$ being the volume of active material.}
	\label{Fig-Structure and FDTD}
\end{figure}

We first consider a three coupled photonic crystal (PhC) nanocavities that supports a non-Hermitian zero mode [see Fig. \ref{Fig-Structure and FDTD}(a)]. The two extreme cavities (sublattice A) are evanescently coupled to the central one (sublattice B). The coupling $K$ can be controlled using the so-called barrier engineering technique~\cite{haddadi2014photonic,hentinger2022direct}, namely, the radius of the barrier holes [yellow holes in Fig.~\ref{Fig-Structure and FDTD}(a)] is modified as $r_b=r_0(1+h)$, where $r_0$ is the radius of the ordinary air holes. Note that such barrier modulation also introduces additional frequency detunings $\Delta\omega_j$. Given the fact that the central cavity is surrounded by two barriers, we can assume that the detuning in the center is twice the one in the two extreme cavities, $\Delta\omega_{1,3}=\Delta\omega, \Delta\omega_2=2\Delta\omega$, the sublattice detuning being $\Delta\omega_2-\Delta\omega_{1,3}=\Delta\omega$. 3D-FDTD simulations, as displayed in Fig.~\ref{Fig-Structure and FDTD}(b), are performed to reveal the mode structure of the optical trimer as a function of the barrier size. To further study the impact of the barrier perturbation $h$ on the coupling $K$ and the sublattice detuning $\Delta\omega$, we carried out another simulation using two coupled PhC cavities and fitted the results with linear Coupled Mode Theory (CMT) [see Fig. \ref{Fig-Structure and FDTD}(c)]. The results can be divided into two different regions: weak cavity-coupling, where $|\Delta\omega(h)|>|K(h)|$ (purple), and strong cavity-coupling, where $|\Delta\omega(h)|<|K(h)|$ (green background). A particularly interesting domain (dashed background) is $-20\%\leq h\leq -10\%$, where the coupling approaches zero ($|K|\approx 0$). In this region, the center cavity is effectively decoupled from the other two and the zero mode becomes one of the eigenmodes of an effective dimer formed by sublattice A; such a parameter regime will be used later on to relate the zero mode frequency to the single cavity one.

We model our system using carrier-dependent Coupled Mode Theory (CD-CMT) \cite{marconi2020mesoscopic,hentinger2022direct}, that governs the time-evolution of coupled complex filed amplitudes $a_j$ in the semiconductor cavities ($j=1,2,3$), in the presence of carrier populations $n_j$,
\begin{gather}
    \label{Eq-sde-field}
    \frac{da_j}{dt}=Ha_j+F_j(t)a_j,\\
    \begin{aligned}
    \label{Hamiltonian}
		H=&\begin{pmatrix}
            \omega_{0}+\alpha g_1+\Delta\omega_1 & K & 0 \\
            K & \omega_{0}+\alpha g_2+\Delta\omega_2 & K\\
            0 & K & \omega_{0}+\alpha g_3+\Delta\omega_3
		\end{pmatrix}\\
		&+i\begin{pmatrix}
            g_1-\frac{1}{\tau} & 0 & 0 \\
            0 & g_2-\frac{1}{\tau} & 0\\
            0 & 0 & g_3-\frac{1}{\tau}
		\end{pmatrix},
	\end{aligned}\\
     \label{Eq-sde-carrier}
    \frac{dn_j}{dt}=P_j-n_j\Gamma_{tot}-\beta\Gamma_{\parallel}(n_j-n_0)|a_j|^2,
\end{gather}
where $\omega_{0}$ and $\tau$ are the resonant frequency and cavity-damping time of a single transparent cavity, repectively, $g_j=\beta \Gamma_\parallel (n_j-n_0)/2$ are the gain rates, $\beta$ is the spontaneous emission factor, $\Gamma_{\parallel}$ is the two-level radiative recombination rate, $\alpha$ is linewidth enhancement ---or Henry--- factor, $n_0$ represents the carrier number at transparency, $P_j$ are the pump rates and $\Gamma_{tot}$ is the total carrier recombination rate. $F_j(t)$ represent Langevin noise terms accounting for spontaneous emission.

We start with a simple case where $\Delta\omega_j=0$, $F_j(t)=0$ and carrier-induced refractive index change effect is neglected ($\alpha=0$). When the two extreme cavities are equally pumped we  set $g_1=g_3=g$ and $g_2=0$. Below the laser threshold ($|a_1|^2=|a_2|^2=|a_3|^2\rightarrow 0$), the eigenvalues read
\begin{gather}
	\epsilon_0=\omega_0+i\left(g-\frac{1}{\tau}\right),\\
	\epsilon_{\pm}=i\left(\frac{g}{2}-\frac{1}{\tau}\right)\pm\frac{1}{2}\sqrt{8K^2-g^2}+\omega_0,
\end{gather} 
where $\epsilon_0$ is the NHPH symmetry protected zero mode with eigenvector being $(1,0,-1)^T$. Hence, the zero mode is a dark one, {\em i.e.} it can only be excited by probing sublattice A; noticeably, there is a $\pi$ phase difference between the two extreme cavities. This $\pi$-phase difference results, in the general non Hermitian case, from twice the $\pi/2$-phase jumps between sublattices as a consequence of the NHPH symmetry~\cite{ge2017symmetry}, and proves to be robust against pump unbalancing. Such a $\pi$-phase difference characterizing the zero mode is a useful fingerprint to identify it experimentally~\cite{hentinger2022direct}.

Figure~\ref{Fig-Structure and FDTD}(d) shows the real and imaginary parts of the eigenvalues of the full ---i.e., including $\alpha$-induced blue-shift effects--- Hamiltonian in Eq.~(\ref{Hamiltonian}). When a single extreme cavity is pumped (blue curves), the highest-frequency mode is more efficiently excited and reach the laser threshold [$\Re (\epsilon)=0$] first. This can be explained by considering the spectral overlap between the excited cavity photons and the hybrid modes; namely, the carrier-induced blue-shift of the excited cavity resonance better overlaps with the blue-detuned hybrid mode ~\cite{hentinger2022direct}. In contrast, the zero mode reaches the threshold before the two others and becomes the lasing mode as long as the two extreme cavities are pumped equally (see red curves).

\begin{figure}[t!]
	\centering
	\includegraphics[width=\linewidth]{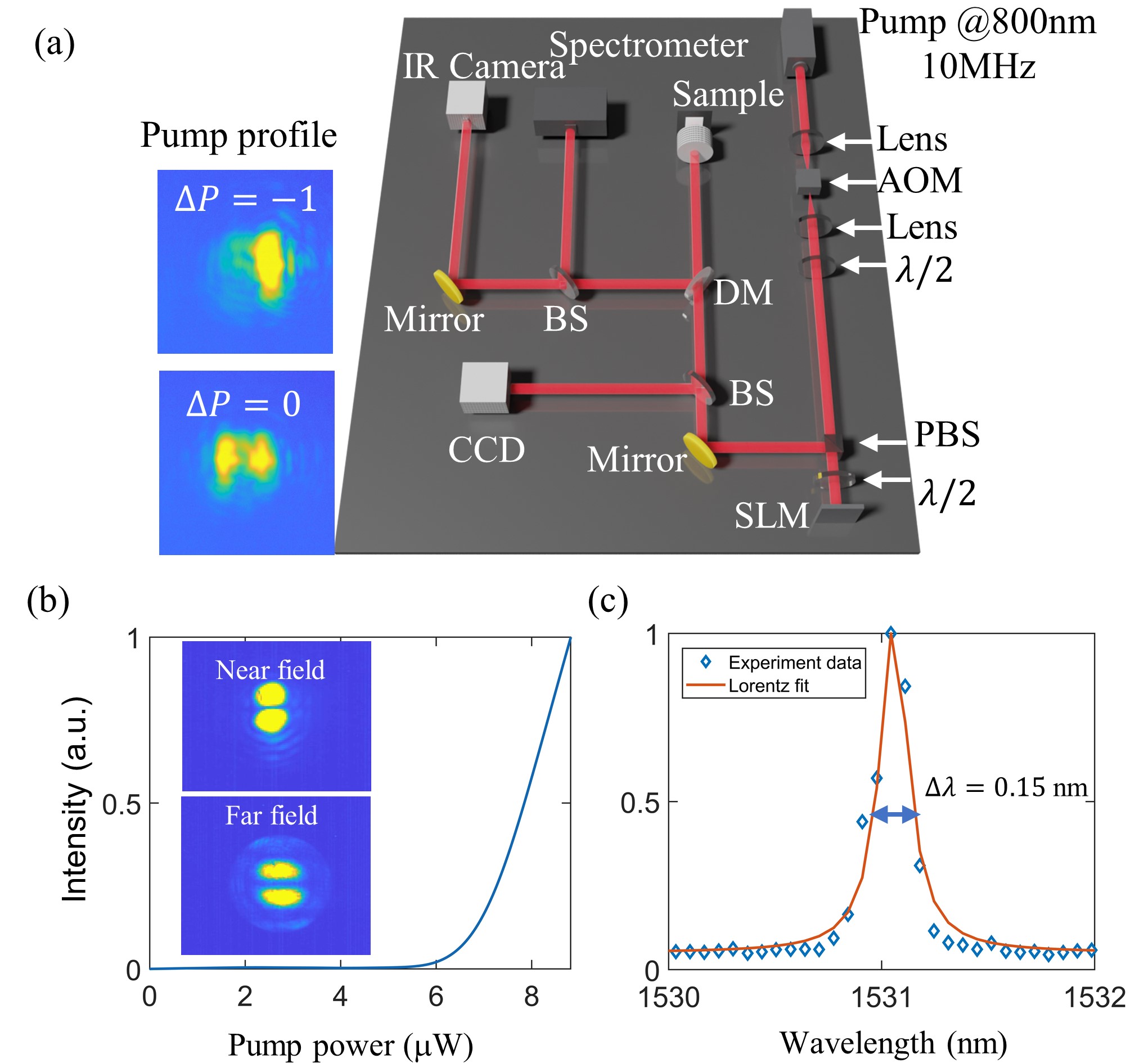}
	\caption{Experimental characterizations of zero mode laser. (a) Experiment setup with the SLM; (b) Light-in/light-out curve of the zero mode laser. The inserts are the near and far fields of the zero mode laser; (c) Experimental measurement and the Lorentz fit of zero mode laser.}
	\label{Fig-first zero mode laser}
\end{figure}

In order to observe the zero mode laser experimentally, we fabricated the three coupled PhC cavities in an Indium Phosphide (InP) suspended membrane, with four embedded InGa$_{0.17}$As$_{0.76}$P QWs~\cite{hamel2015spontaneous}. We performed photoluminescence (PL) experiments to study the emission properties of the zero mode. We choose a pulsed pump-laser (100ps duration and 10 MHz repetition rate) as the pump source in order to reduce thermal effects. As we discussed before, the zero mode cannot be effectively excited using only one spot, hence, we employ a spatial light modulator (SLM) to reshape the pump configuration. Our liquid crystal-based SLM is operated in amplitude modulation mode [Fig. \ref{Fig-first zero mode laser}(a)]. This configuration requires two half-wave plates, the first one (close to the AOM) is used to increase the global intensity of the pattern, and the second one (close to the SLM), which is rotated to be $22.5^\circ$ with respect to the horizontal plane, maximizes the contrast between the pump pattern and the background. Such a modulation allows us to control the pump intensities  in three cavities independently. Two typical pump patterns are depicted in the insets of  Fig.~\ref{Fig-Structure and FDTD}(a). The modified pump laser is then focused down on the sample through a microscope objective (100$\times$ IR with 0.95 NA) and the radiated PL is spectrally resolved with a spectrometer.

We characterize the nanolaser emission under nearly equal pumping of the two extreme cavities. The light-in/light-out emission [Fig. \ref{Fig-first zero mode laser}(b)] together with the emission linewidth [Fig. \ref{Fig-first zero mode laser}(c)] are strong evidences of a zero mode laser obtained in the strong intercavity coupling region ($h=0\%$, for which $|K|\gg |\Delta \omega|$), at $\lambda\approx1531.11$nm. In Fig. \ref{Fig-first zero mode laser}(b) a clear laser threshold is observed for an average pump power of about $6\,\mu$W. To verify that such a mode is indeed a zero mode, we measure its near and far field patterns using an InGaAs infrared camera. In the near field, no energy is detected in the center; in addition, a node is  observed in the far field image, revealing an anti-symmetric field distribution [insets in Fig.~\ref{Fig-first zero mode laser}(b)]. As a matter of fact, a $\pi$ phase-difference between the two extreme cavities is compatible with the predicted $\pi/2$ phase-jumps between adjacent cavities belonging to different sublattices of a NH zero mode \cite{ge2017symmetry}.

\begin{figure}[t!]
	\centering
	\includegraphics[width=\linewidth]{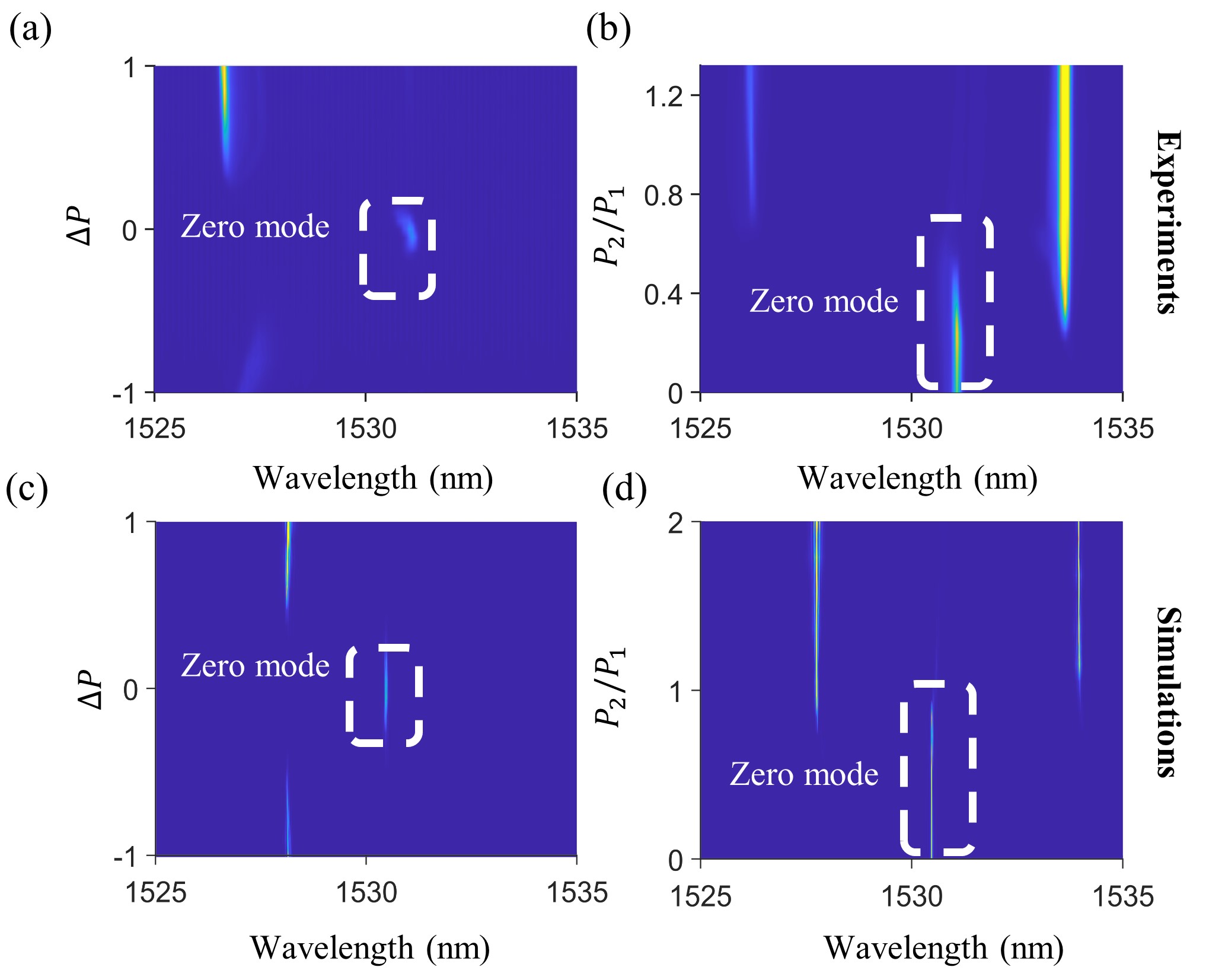}
	\caption{Experimental observation of zero mode laser in the strong coupling region ($h=0\%$ and lattice constant is $a=408$nm). (a) PL map when the pumps in the two extreme cavities are unbalanced; (b) PL map when the pump in the center cavity is increased. The dashed boxes indicate the location of zero mode laser.(c) and (d) are the simulations of (a) and (b), respectively. (e) the experimental measurement and the Lorentz fit of zero mode laser; (f) light-in/light-out curve of the zero mode laser; Here in the simulations we use:  $K=12.5/\tau$ , $\Delta\omega_1=\Delta\omega_3=-3.38/\tau, \Delta\omega_2=2\Delta\omega_{1,3}, \omega_0=195.81$THz, $P_1+P_2=3P_0$, with $P_0$ being the threshold for the single cavity. The rests are the same as the parameters in Fig.~\ref{Fig-Structure and FDTD}.}
	\label{Fig-robustness of zero mode laser}
\end{figure}

With the aim of further testing the robustness of the zero mode laser, we define a new control parameter $\Delta P=(P_3-P_1)/(P_1+P_3)$ that conserves the total pump power ($P_1+P_3=const$). Panel (a) in Fig. \ref{Fig-robustness of zero mode laser} shows the spectral intensity as a function of both wavelength and $\Delta P$. Note that only one of the extreme cavities is pumped when $\Delta P=\pm 1$, and two of them are pumped equally when $\Delta P=0$. The zero mode laser is efficiently excited in the range $|\Delta P|\leq 0.15$. Figure \ref{Fig-robustness of zero mode laser}(c) depicts the numerical calculation results using Eqs. (\ref{Eq-sde-field})-(\ref{Eq-sde-carrier}), showing very good agreement with the experiment.

Interestingly, we can control the zero mode laser by injecting power in the center cavity ($P_1=P_3$, $P_2=0\rightarrow 1.2P_{1}$); the experimental results are displayed in Fig.~\ref{Fig-robustness of zero mode laser}(b). Figure.~\ref{Fig-robustness of zero mode laser} (d) is the corresponding simulation results, which are in good agreement with the experiment. We observe that the zero persists as long as the center cavity is absorptive, namely, $g_2\leq 0$, which takes place for $P_2/P_1<0.5$; yet, at the onset  of net gain in the center cavity ($g_2>0$), the zero mode switches off and the energy is transferred to side modes. This result implies that the visibility of the zero mode laser can be dynamically controlled by vaying the pump power in the center cavity.

A most important property of a lasing zero mode is that its frequency should correspond to the single cavity one, in this case modified by the presence of a barrier, $\omega_0+\Delta\omega$. However, it is difficult to directly compare the zero mode with the single cavity nanolaser experimentally because the latter would belong to a different lithographic realization, therefore it is likely to be detuned with respect to the coupled cavity central frequency. In order to cope with this, we move to a weaker coupling regime in which we can effectively decouple the system by unbalancing the optical pumping. 

As predicted in Fig.~\ref{Fig-Structure and FDTD}(\ag{b}), the coupling is near zero in the range of $-20\%\leq h \leq -10\%$, the center cavity {(sublattice B)} becomes decoupled from sublattice A and therefore the system can be treated as an effective optical dimer given by the two coupled extreme cavities~\cite{hentinger2022direct}. In order to compute the effective coupling parameters we apply the Schrieffer-Wolff transformation~\cite{schrieffer1966relation,zhong2021control} to the Hamiltonian in Eq.~(\ref{Hamiltonian}) (here, for the sake of simplicity, we take $\alpha=0$ and only consider Hermitian terms),
\begin{gather}
	\begin{split}
		\label{eq4}
		H_{e}&=\hat{U}^\dagger H\hat{U}=\\
		&\begin{pmatrix}
			\omega_0+\Delta\omega_{e} & K_{e} & 0\\
			K_{e} & \omega_0+\Delta\omega_{e} & 0\\
			0 & 0 & \omega_0+\Delta\omega_{e}'
		\end{pmatrix},
	\end{split}
\end{gather}
with the eigenvectors being $[a_1,a_3,a_2]^T$. Here $\hat{U}$ is a unitary operation used to decouple the center cavity from the rest of the system. The effective coupling strength reads $K_e=K^2/\Delta\omega$, and the effective detunings for the extreme and center cavities are $\Delta\omega_{e}=K^2/\Delta\omega+\Delta\omega$ and  $\Delta\omega_{e}'=2K^2/\Delta\omega+2\Delta\omega$, respectively. The eigenvalues of Eq.~(\ref{eq4}) are
\begin{gather}
	\label{eq5}
	\omega=\Delta\omega+\omega_{0},\\
	\label{eq6}
	\omega=\Delta\omega +\frac{2K^2}{\Delta\omega}+\omega_0,
\end{gather}
{Equations} (\ref{eq5}) and (\ref{eq6}) are the hybrid frequencies of the effective dimer. Eq.(\ref{eq5}) implies that the effective zero mode in the dimer is lasing at the frequency of the {uncoupled} extreme cavity. {Note that,} unlike the conventional two coupled cavities, where the eigenfrequencies are split symmetrically with respect to the {uncoupled cavity} frequency, here the two eigenvalues of the effective dimer are not equally distant from the single cavity frequency. 

Table I displays the parameters of the effective dimer compared with the actual ones for the two coupled cavity case.
\begin{table}
\begin{center}
	\begin{tabular}{|c|c|c|c|c|}
		\hline
		$h$ & -20\% &	-15\% & -10\% & -5\%\\
		\hline
		$K$(actual two cavities) & -7.44 & -0.84 & 4.57 & 8.96\\
		\hline
		$K_e$ & 1.08 & 0.018 & 0.80 & 5.51\\
		\hline
	\end{tabular}
\end{center}
\caption{Comparison of the couplings between actual (above) and the effective two coupled cavities (below). Parameters are normalized to the single cavity loss rate $1/\tau$.}
\end{table}
Clearly, the effective coupling is smaller compared with the actual system. Thus, unlike Figs \ref{Fig-robustness of zero mode laser}(a) and (c), unbalancing the pumping may effectively decouple the extreme cavities, and therefore the single cavity frequency can be directly compared with that of the zero mode.

\begin{figure}[t!]
	\centering
	\includegraphics[width=\linewidth]{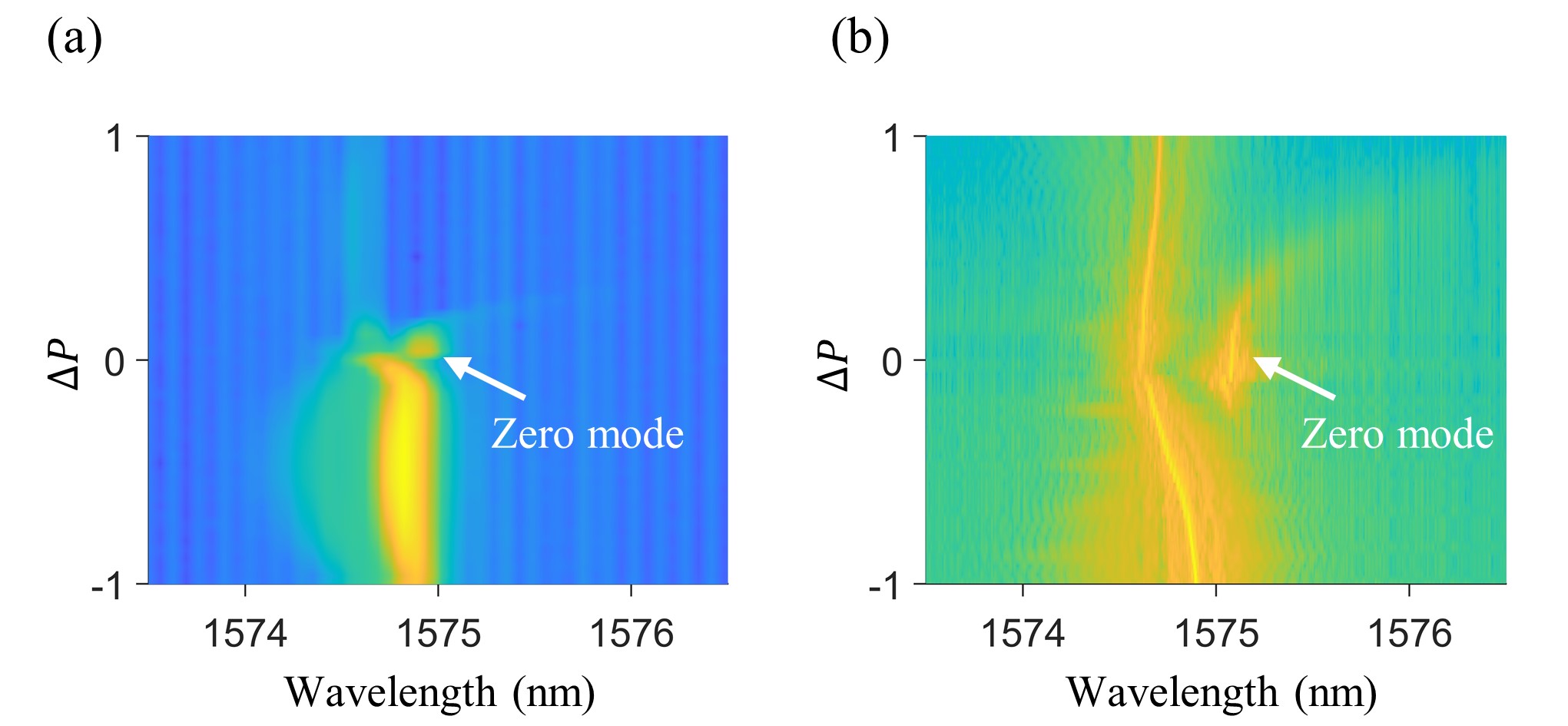}
	\caption{Experimental observation of zero mode laser in the weak coupling region ($h=-20\%, a=422$nm). (a) experimental comparison between the zero mode laser and the single cavity, the PL map is plotted in logarithmic scale; (b) simulation result using the effective two coupled system model. Here the parameters are: $K=1.08/\tau, \Delta\omega_1=0, \Delta\omega_2=1.2g$, and $\omega_0=190.28$THz.}
	\label{fig.3}
\end{figure}

In the following we measure the PL maps in the near zero coupling region. We point out that the zero mode is only warranted for nonzero coupling, therefore vanishing coupling, predicted for $h=-15\%$ (Fig. \ref{Fig-Structure and FDTD}), must be avoided; we then choose, for this experiment, $h=-20\%$. 
The results are displayed in Fig.~\ref{fig.3}(a). As the unbalanced pump parameter $\Delta P$ decreases from $\Delta P=0$ to $\Delta P=-1$, the leftmost mode red shifts and its wavelength approaches to that of the zero mode. This observation implies that the zero mode frequency is close to the one corresponding to a single, uncoupled cavity, albeit perturbed by the presence of the barrier ($\omega_0+\Delta\omega$). 
Figure.~\ref{fig.3}(b) {shows} the simulation results of the effective dimmer model, which agrees with the experimental result nicely.

In conclusion, we have experimentally demonstrated a lasing zero mode warranted by the non-Hermitian particle-hole (NHPH) symmetry of a three coupled photonic crystal nanocavity array. Such a symmetry protected mode can be efficiently excited by pumping the two extreme cavities with equal power, but the zero mode survives within a range of pump power unbalance of $15\%$. Furthermore, the intensity of the zero mode can be dynamically controlled by pumping the center cavity, which eventually transfers the energy to side modes; noteworthy, we have observed that the zero mode is robust upon increase of the central cavity pump power up to $50\%$ of the extreme cavity pump level. We have also shown that the three coupled cavities can be reduced to a dimmer system when the coupling is weak compared to the sublattice detuning, which is used to actively decouple the system and confirms that the zero mode oscillates at the frequency of a single cavity. Our work provides a flexible way to excite and control the lasing zero mode via SLM pump patterning, whose manipulation and symmetry protection features may be of interest in potential applications such as optical computing\cite{tirabassi2022binary} and fault-tolerant quantum computation \cite{levene2003zero,fu2008superconducting}.

\section*{acknowledgments}
This work is partially supported by the French National Research Agency (ANR), Grants No. ANR UNIQ DS078 and ANR-22-CE24-0012-01, the European Union in the form of Marie Sk\l odowska-Curie Action grant MSCA-841351, and by the RENATECH network. K. J. acknowledges the finial support for China Scholarship Council (No.202006970015)

\nocite{*}

\bibliography{main}

\end{document}